\documentclass[structabstract]{aa}
\usepackage{amsmath}
\usepackage{graphicx}
\usepackage{txfonts}
\usepackage{epsfig}
\nonstopmode

\begin{document}

   \title{Mass retention efficiencies of He accretion onto carbon-oxygen white dwarfs and type Ia supernovae}

   \author{C. Wu  \inst{1,2,3,4}
          \and
          B. Wang  \inst{1,2,3,4}
          \and
          D. Liu \inst{1,2,3,4}
          \and
          Z. Han \inst{1,2,3,4}
          }
   \institute{Yunnan Observatories, Chinese Academy of Sciences, Kunming 650216, China;
             \email{wcy@ynao.ac.cn; wangbo@ynao.ac.cn}
             \and
             Key Laboratory for the Structure and Evolution of Celestial Objects, Chinese Academy of Sciences, Kunming 650216, China
             \and
             University of Chinese Academy of Sciences, Beijing 100049, China
             \and
             Center for Astronomical Mega-Science, Chinese Academy of Sciences, 20A Datun Road, Chaoyang District, Beijing, 100012, China
              }

\abstract
{Type Ia supernovae (SNe Ia) play a crucial role in studying cosmology and galactic chemical evolution. They are thought to be thermonuclear explosions of carbon-oxygen white dwarfs (CO WDs)
when their masses reach the Chandrasekar mass limit in binaries. Previous studies have suggested that He novae may be progenitor candidates of SNe Ia. However, the mass
retention efficiencies during He nova outbursts are still uncertain.}
{In this article, we aim to study the mass retention efficiencies of He nova outbursts and to investigate whether SNe Ia can be produced through
He nova outbursts.}
{Using the stellar evolution code {\em Modules for Experiments in Stellar Astrophysics}, we simulated a series of multicycle He-layer flashes, in which the initial WD masses range from $0.7$ to $1.35\,{M}_\odot$ with various
accretion rates.}
{We obtained the mass retention efficiencies of He nova outbursts for various initial WD masses, which can be used in the binary population synthesis studies.
In our simulations, He nova outbursts can increase the mass of the WD to the Chandrasekar mass limit and the explosive carbon burning can be triggered in the center of the WD; this suggests that He nova outbursts can produce SNe Ia. Meanwhile, the mass retention efficiencies in the present work are lower than those of previous studies, which leads to a lower birthrates of SNe Ia through the WD + He star channel. Furthermore, we obtained the elemental abundances distribution at the moment of explosive carbon burning,
which can be used as the initial input parameters in studying explosion models of SNe Ia.}
{}


\titlerunning{Mass retention efficiencies of He accretion onto carbon-oxygen white dwarfs and SNe Ia}

\authorrunning{C. Wu et al.}

   \maketitle


\section{Introduction}

Type Ia supernovae (SNe Ia) are successful cosmological distance indicators because of their uniform luminosities, which reveal the accelerating expansion of the Universe. It is widely accepted that SNe Ia originate from thermonuclear explosions of carbon-oxygen white dwarfs (CO WDs) (e.g., Hoyle \& Fowler 1960). However, their progenitor models and explosion models are not completely understood (e.g., Nomoto et al. 1997; Wang \& Han 2012). Two progenitor models have been discussed for several decades. One is the single-degenerate (SD) model in which a CO WD accretes material from its non-degenerate companion (e.g., a main-sequence star, a red giant star, or a He star; see Whelan \& Iben 1973; Nomoto et al. 1984; Li \& van den Heuvel 1997; Han \& Podsiadlowski 2004). The  other is the double-degenerate (DD) model in which two CO WDs merge owing to the loss of orbital angular momentum driven by gravitational wave radiation (e.g., Iben \& Tutukov 1984; Webbink 1984). At present, several explosion models have been proposed, such as the pure deflagration model (e.g., Nomoto et al. 1984), delayed detonation model (e.g., Khokhlov 1991; H\"oflich et al. 1995), and double detonation model (e.g., Taam 1980; Nomoto 1982a).

In the SD model, the WD + He star channel makes a main contribution to the birthrate of young SNe Ia (e.g., Wang et al. 2009a,b; Ruiter et al. 2009). In this channel, the accreted He-rich material from the donor is burned into carbon and oxygen, leading to the mass growth of the WD. The WD may produce an SN Ia when its mass approaches the Chandrasekar mass (Ch-mass) limit. Different accretion rates may have influence on the final fate of the accreting WD (e.g., Nomoto 1982b; Piersanti et al. 2014; Wang et al. 2015). At high accretion rates, the accreted material may pile up on the surface of the WD and the WD may evolve into a He-giant-like star (e.g., Nomoto 1982b); at low accretion rates successive He-layer flashes may occur owing to the unsteady He burning (e.g., Piersanti et al. 2014; Hillman et al. 2016). The He-rich material can only be burned into carbon and oxygen steadily when the accretion rate is in a narrow regimes (e.g., Nomoto 1982b; Wang et al. 2015). In the observations, the process of He-layer flashes is related to some objects, such as He novae and AM CVn stars (e.g., Nelemans et al. 2001; Bildsten et al. 2007; Shen \& Bildsten 2009; Piersanti et al. 2015). V445 Pup is the first identified He nova although it is still not well studied (e.g., Ashok \& Banerjee 2003; Woudt et al. 2009). Kato et al. (2008) inferred that the mass of the WD in V445 Pup is more than $1.35{M}_\odot$ based on the light curve fitting in their study and they suggested that V445 Pup is a strong progenitor candidate of SN Ia as the WD can increase its mass in He-layer flashes (see also Woudt et al. 2009).

The mass retention efficiencies of He-rich material play a key role in the binary population synthesis (BPS) studies of the SD channel (e.g., Wang et al. 2009b, 2013). Although the retention efficiencies have long been studied, they are still not completely understood (e.g., Piersanti et al. 2014; Hillman et al. 2016; Wang et al. 2015). Kato \& Hachisu (2004; hereafter KH04) studied the mass retention efficiencies of CO WDs that undergo only one He-layer flash based on an optically thick wind assumption (e.g., Hachisu et al. 1996). However, the mass retention efficiencies and final outcomes of WDs may be changed after multicycle He nova outbursts (e.g., Piersanti et al. 2014; Wang et al. 2015). Meanwhile, the optically thick wind assumption sensitively depends on the metallicity, which does not work at low metallicity (e.g., Kobayashi et al. 1998). The observed SNe Ia at high redshifts cannot be explained by the optically thick wind assumption (e.g., Jones et al. 2013).

In this article, we aim to investigate the mass retention efficiencies of CO WDs after multicycle He nova outbursts and to study its final outcomes. In Sect. 2, we introduce basic assumptions and methods for our numerical calculations. The calculation results are given in Sect. 3. In Sect. 4, we compare the mass retention efficiencies with previous studies and provide the SN Ia birthrates of WD + He star channel. Finally, we present the discussions and conclusions in Sect. 5.

\section{Methods}

The stellar evolution code used in our simulations is \texttt{Modules for Experiments in Stellar Astrophysics} (MESA; see Paxton et al. 2011, 2013, 2015) [version, 7624]. The OPAL opacity is adopted (e.g., Igleslas \& Rooers, 1996), which is applicable to extra carbon and oxygen during the He burning. We adopted \texttt{co\_burn.net} as the nuclear reaction network in our simulations. This nuclear reaction network includes the isotopes needed for He, carbon, and oxygen burning (i.e., $^{\rm 3}{\rm He}$, $^{\rm 4}{\rm He}$, $^{\rm 7}{\rm Li}$, $^{\rm 7}{\rm Be}$, $^{\rm 8}{\rm B}$, $^{\rm 12}{\rm C}$, $^{\rm 14}{\rm N}$, $^{\rm 15}{\rm N}$, $^{\rm 16}{\rm O}$, $^{\rm 19}{\rm F}$, $^{\rm 20}{\rm Ne}$, $^{\rm 23}{\rm Na}$, $^{\rm 24}{\rm Mg}$, $^{\rm 27}{\rm Al}$, and $^{\rm 28}{\rm Si}$); these are coupled by 57 reactions.

Our calculations involve two suite cases in MESA, \texttt{make\_co\_wd}, and \texttt{wd2}. The suite case \texttt{make\_co\_wd} is used to construct initial CO WD models from main-sequence stars. Firstly, we evolve a number of main-sequence stars to the asymptotic giant branch and then remove their outer envelopes, leaving a series of pure CO cores. After these hot CO cores evolve to the WD cooling phases, we choose the CO WDs with masses from $0.7$ to $1.35\,{M}_\odot$ and central temperatures from $6.9\times10^7$ to $2.07\times10^8{\rm K}$ as our initial WD models. The suite case \texttt{wd2} is used to calculate the process of mass accretion onto WDs, which includes some options that can control the mass gain and loss processes for nova outbursts. We set the He mass fraction and metallicity of accreted material to be 0.98 and 0.02, respectively. The accretion rates range from $4\times10^{-8}$ to $1.6\times10^{-6}\,{M}_\odot\,\mbox{yr}^{-1}$, in which these WDs undergo multicycle He-layer flashes. In our simulations, the accretion rates are assumed to remain constant. Although the mass transfer rates will change during the binary evolution, we only focus on the evolution of WDs for various accretion rates rather than binary systems.

We assume super-Eddington wind as the mass-loss mechanism during He-layer flashes (e.g., Denissenkov et al. 2013, 2017). If the luminosity on the surface of the WD ($L_{\rm eff}$) exceeds the Eddington luminosity ($L_{\rm Edd}$), under our assumptions, super-Eddington wind would be triggered and blow away part of the accumulated material outside the CO core. The value $L_{\rm Edd}$ can be expressed as
\begin{equation}
{L}_{\rm Edd}=\frac{4\pi{\rm G}{\rm c}M_{\rm WD}}{\kappa},
\end{equation}
where ${\rm G}$, ${\rm c}$, ${M}_{\rm WD}$, and $\kappa$ are the gravitational constant, vacuum speed of light, WD mass, and Rosseland mean opacity on the surface of the WD, respectively. The wind mass-loss rate during the process of nova outbursts can be expressed as follows:
\begin{equation}
\dot{M}=\frac{2\eta_{\rm Edd}(L_{\rm eff}-L_{\rm Edd})}{v^{2}_{\rm esc}},
\end{equation}
where $\eta_{\rm Edd}$ is the super-Eddington factor and $v^{2}_{\rm esc}=\frac{2GM_{\rm WD}}{R_{\rm WD}}$, in which $v_{\rm esc}$ and $R_{\rm WD}$ are the escape velocity and radius of the WD, respectively. In the mass-loss process, the wind is assumed to have the velocity of $v_{\rm esc}$, which means that the wind has zero kinetic energy at infinity. The value $\eta_{\rm Edd}$ in this equation is a parameter that is used to control the efficiency of the wind mass loss and it can range from $0$ to $1.0$. In our simulations, when the super-Eddington wind is triggered, we assume that all energy above the Eddington luminosity is used to eject mass (i.e., $\eta_{\rm Edd}=1$).

\section{Results}

\subsection{An example of He-layer flashes}

\begin{figure}[]
\begin{center}
\includegraphics[width=9.5cm,angle=0]{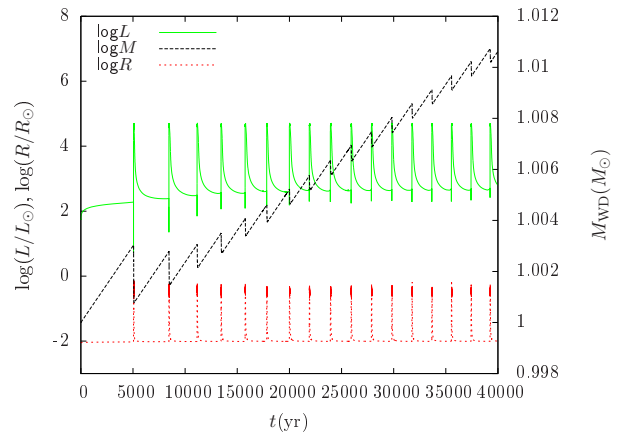}
 \caption{Long-term evolution of a $1.0{M}_\odot$ accreting WD undergoes multicycle nova outbursts. The evolution of the luminosity, radius, and mass changing with time are shown in this figure in which $\dot{M}_{\rm acc}=6\times 10^{-7}\,{M}_\odot\,\mbox{yr}^{-1}$.}
  \end{center}
\end{figure}

Fig.\,1 shows the evolution of luminosity, mass, and radius as functions of time of a $1.0{M}_\odot$ CO WD during multicycle He-layer flashes in $4\times 10^{4}$ years, where we set the mass transfer rate $\dot{M}_{\rm acc}$ as $6\times 10^{-7}\,{M}_\odot\,\mbox{yr}^{-1}$ and the super-Eddington wind factor $\eta_{\rm Edd}$ as $1$. During the mass accumulation process, super-Eddington wind can be triggered once the luminosity exceeds ${L}_{\rm Edd}$. Part of the material is blown away by the wind and the WD mass decreases rapidly. The mass of He-layer material blown away by the wind decreases after every nova outburst, since the successive He-layer flash heats the CO core, leading to the degeneracy of the accreted material decreases; this causes the decrease in energy released by He burning. Meanwhile, the WD has grown in mass for about $0.0105{M}_\odot$ in $4\times 10^{4}$ years, which indicates that the mass retention efficiency approximates to $44\%$ in this example.

\begin{figure}[]
\begin{center}
\includegraphics[width=9.5cm,angle=0]{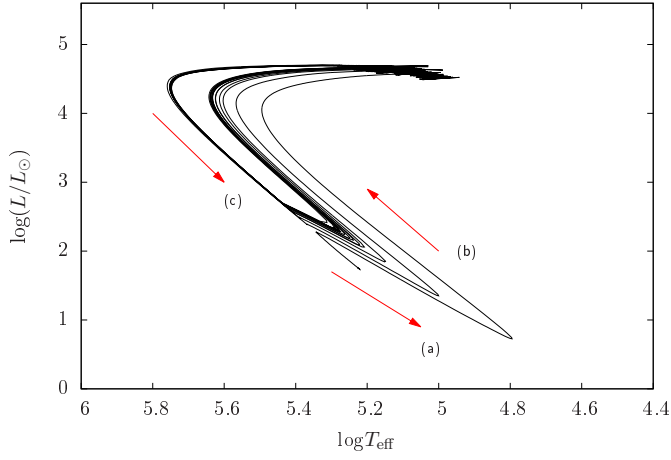}
 \caption{Similar to Fig.\,1, but for the Hertzsprung-Russell diagram. Different phases in every nova cycle are labeled in this figure.}
  \end{center}
\end{figure}

In Fig.\,2, we present the Hertzsprung-Russell diagram of the $1.0{M}_\odot$ CO WD that undergoes successive nova outbursts. Every nova cycle can be divided into three phases as follows: (a) He-rich material piles up on the surface of the WD constantly and the surface temperature $T_{\rm eff}$ and luminosity $L$ decrease owing to the expansion of He layer caused by He burning. (b) Thermal energy is transformed outside the layer via convection in the burning zone, resulting in $L_{\rm eff}$ increasing rapidly until it exceeds $L_{\rm Edd}$ and triggers the super-Eddington wind. (c) The WD continues to accrete material after the wind stops and proceeds to the cooling phase, leading to the decrease of $L_{\rm eff}$ and $T_{\rm eff}$. The minimum of $L_{\rm eff}$ and $T_{\rm eff}$ increase in every nova outburst, since the CO core is heated by the successive He-layer burning.

\subsection{Mass retention efficiencies and nova cycle durations}

\begin{figure}[]
\begin{center}
\includegraphics[width=9.5cm,angle=0]{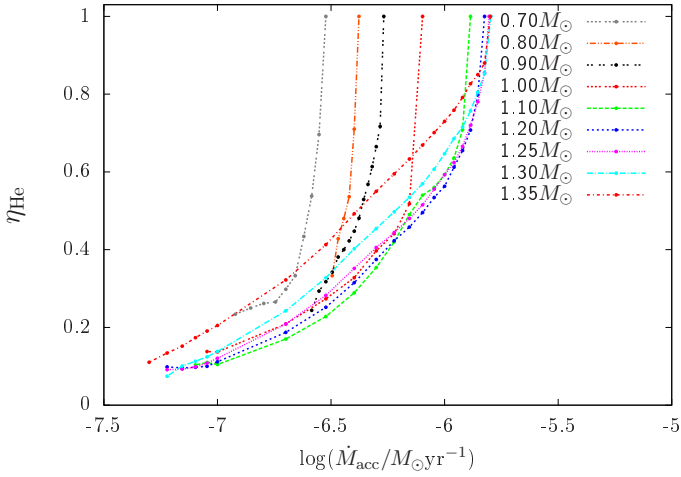}
 \caption{Mass retention efficiencies ($\eta_{\rm He}$) of various initial WD masses in a wide range of accretion rates.}
  \end{center}
\end{figure}

The mass retention efficiencies ($\eta_{\rm He}$) of He nova outbursts are important parameters for progenitors of SNe Ia (e.g., Brooks et al. 2016), which are defined by the ratio of mass accumulation to the evolution time. Fig.\,3 shows $\eta_{\rm He}$ changing with accretion rates for various initial WD masses. From this figure, we can see that $\eta_{\rm He}$ increases with accretion rate, since the thermonuclear runaway of the He layer is weaker for high accretion rates, resulting in less material blown away by the super-Eddington wind. The He-layer flashes becomes too weak to eject layer material, if the accretion rate is higher than a critical value (i.e., the lower limit of the steady burning regime), which means that all the accreted material is accumulated on the WD (i.e., $\eta_{\rm He}=1$). Moreover, compared with low-mass WDs, massive WDs have stronger surface gravitational acceleration and higher central temperature, resulting in more material accumulating on the massive WD. Thus, $\eta_{\rm He}$ is higher for the massive WD at a given accretion rate. Polynomial fitting formulae of $\eta_{\rm He}$ are provided in Appendix A.

\begin{figure}[]
\begin{center}
\includegraphics[width=9.5cm,angle=0]{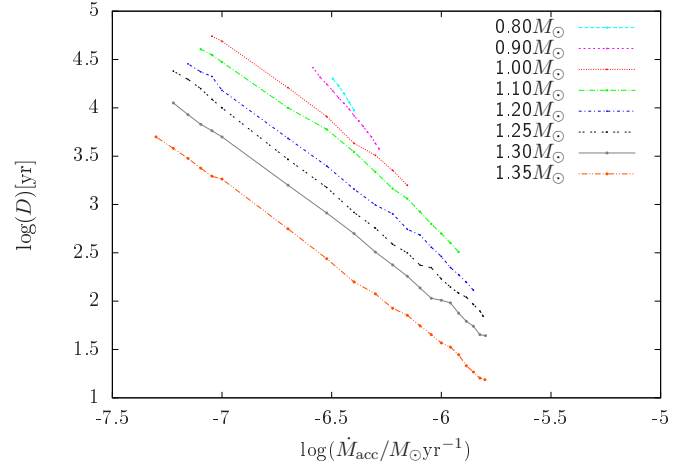}
 \caption{Nova cycle durations ($D$) vs. accretion rates for various WD masses.}
  \end{center}
\end{figure}

Nova cycle duration is the time interval between two successive He-layer flashes, which is an important feature for nova outbursts (e.g., Hillman et al. 2015, 2016). Fig.\,4 shows nova cycle durations changing with the accretion rates for various initial WDs. The durations become shorter for massive initial WDs at a given accretion rate. This is because the He-layer masses needed for He burning are smaller for massive WDs owing to their higher initial central temperatures and stronger surface gravity accelerations. For a given WD, He-layer flashes can occur when the accumulated He-layer mass reaches the same critical value, resulting in that the durations become shorter when the accretion rates increase. Thus, the recurrent phenomena of He-layer flashes on the massive WDs with higher accretion rates would occur more frequently than those of low-mass WDs with lower accretion rates. Linear fitting formulae of durations are provided in Appendix B.

\subsection{SNe Ia from He nova outbursts}

\begin{figure}[]
\begin{center}
\includegraphics[width=9.5cm,angle=0]{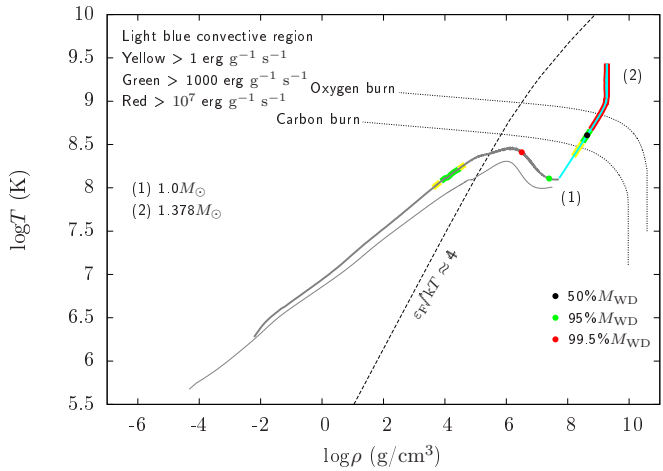}
 \caption{Density and temperature profiles at initial and final moments of the CO WD during He-layer flashes, which presented by lines (1) and (2), respectively. Here, the initial WD mass and accretion rate are $1.0{M}_\odot$ and $7.5\times10^{-7}\,{M}_\odot\,\mbox{yr}^{-1}$, respectively.}
  \end{center}
\end{figure}

If the He nova outbursts continue to occur, the WD mass may eventually reach the Ch-mass limit, and the explosive carbon burning is ignited in the center of the WD, resulting in SN Ia. We assume that the starting point of the central temperature dramatically increasing as the criterion of explosive carbon ignition in our simulations (see also Chen et al. 2014; Wu et al. 2016). Fig.\,5 shows the density and temperature profile of the WD, where lines (1) and (2) represent the initial and final structures of the WD during the long-term evolution, respectively. We start the simulation with a $1.0{M}_\odot$ WD and the accretion rate is set to be $7.5\times10^{-7}\,{M}_\odot\,\mbox{yr}^{-1}$. After about $8.5\times10^{5}$ years in evolution time, the WD increases its mass to $1.378{M}_\odot$. The central temperature at this moment increases sharply, passes the explosive carbon ignition point, and eventually reaches $2.1\times10^{9}{\rm K}$. The maximum nuclear reaction rate in the WD center released by the explosive carbon burning is about $7.8\times10^{24}\,{\rm erg}\,\mbox{g}^{-1}\,\mbox{s}^{-1}$ and over $90\%$ of the CO core is in the convection burning zone. This result indicates that an SN Ia can be produced through He nova outbursts.

\begin{figure}[]
\begin{center}
\includegraphics[width=9.5cm,angle=0]{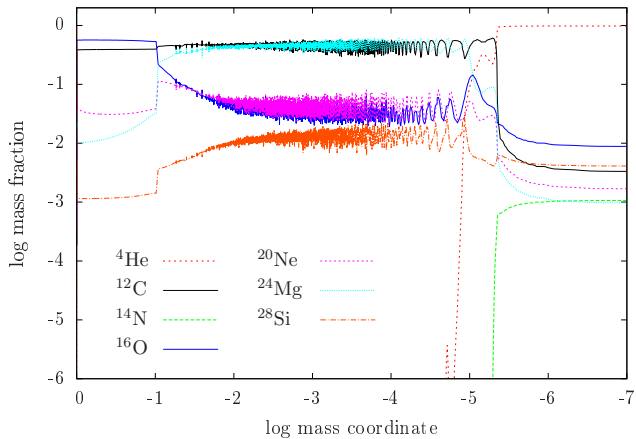}
 \caption{Elemental abundance profiles at the moment of central explosive carbon burning, where the initial WD mass and accretion rate are $1.0{M}_\odot$ and $7.5\times10^{-7}\,{M}_\odot\,\mbox{yr}^{-1}$, respectively.}
  \end{center}
\end{figure}

The elemental abundances distribution at the moment of explosive carbon burning plays a crucial role in studying explosion models of SNe Ia. In Fig.\,6, we present the distribution of main elemental abundances during the explosive carbon burning. As the burning wave propagates from the inside out, carbon in the center of the WD is transformed into neon, which results in the abundance of neon increases about an order of magnitude. During the nova outbursts, He-layer burning transforms the He into carbon, oxygen, and some other intermediate mass elements. These elements are accumulated onto the surface of the WD, leading to the elemental abundances fluctuating on the surface of the CO core. These results can be used as initial input parameters for explosion models of SNe Ia.

\section{Comparison with previous studies}

\begin{figure}[]
\begin{center}
\includegraphics[width=9.5cm,angle=0]{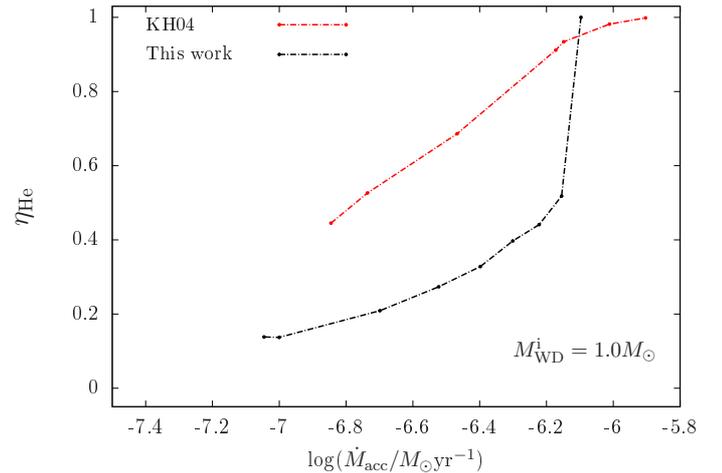}
 \caption{Mass retention efficiency ($\eta_{\rm He}$) of a $1.0{M}_\odot$ WD as a function of mass accretion rate. The black dotted line is the result of our work, whereas the red dotted line is taken from KH04.}
  \end{center}
\end{figure}

KH04 investigated mass retention efficiencies changing with He accretion rates and initial WD masses under the optically thick wind assumption. In Fig.\,7, we compare $\eta_{\rm He}$ of a $1.0{M}_\odot$ WD with that of KH04. This figure implies that $\eta_{\rm He}$ given in our studies is obviously lower. Although a different wind assumption may have influence on the wind mass-loss process, the obvious difference is mainly because KH04 only simulated the He-layer flash one time. The values of $\eta_{\rm He}$ are more reliable if the WDs undergo multicycle He nova outbursts.

\begin{figure}[]
\begin{center}
\includegraphics[width=9.5cm,angle=0]{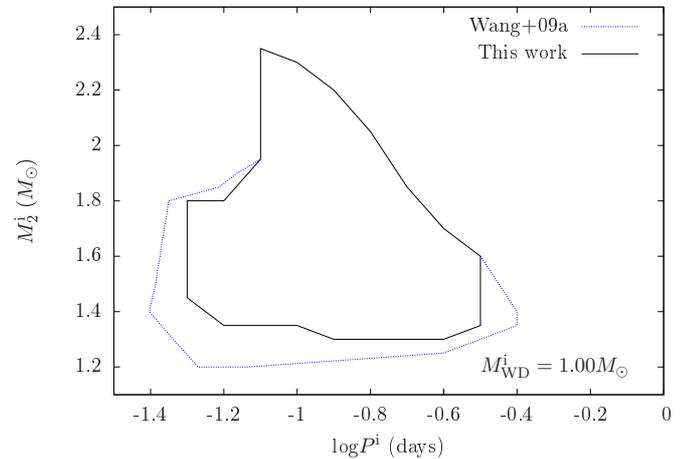}
 \caption{Region in the orbital period-companion mass plane for WD binaries that can produce SNe Ia for initial WD mass of $1.0{M}_\odot$. The black solid line is the region for WD binaries that can produce SNe Ia in our work, whereas the blue dotted line is taken from Wang et al. (2009a).}
  \end{center}
\end{figure}

\begin{figure}[]
\begin{center}
\includegraphics[width=9.5cm,angle=0]{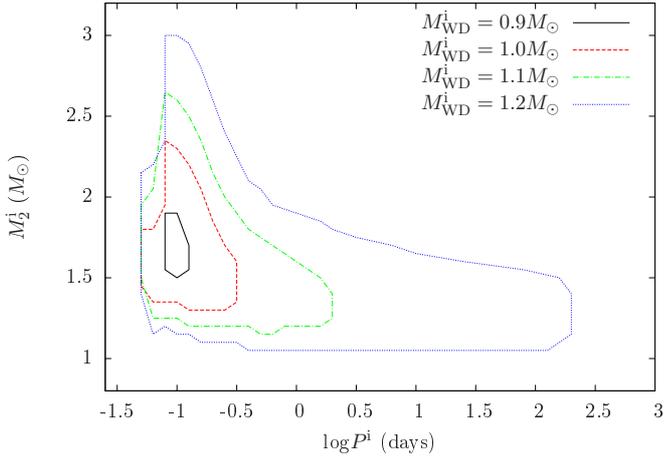}
 \caption{Regions in the orbital period-companion mass plane for WD binaries that can produce SNe Ia for initial WD masses of 0.9, 1.0, 1.1, and $1.2{M}_\odot$.}
  \end{center}
\end{figure}

Wang et al. (2009a) calculated the binary evolution of WD + He star systems by adopting the prescription of KH04 as the mass retention efficiencies of the He-layer flashes onto the WDs, and they obtained the parameter spaces for producing SNe Ia from the WD + He star channel. However, the results may be different if $\eta_{\rm He}$ becomes lower. In order to analyze how $\eta_{\rm He}$ influence the parameter spaces of SNe Ia, we employed the Eggleton stellar evolution code (see Eggleton 1973) to calculate the binary evolution of WD + He star systems. The basic assumptions and methods in these simulations are similar to Wang et al. (2009a). The region in the black solid grid of Fig.\,8 is the parameter space of binary systems for the initial WD mass of $1.0{M}_\odot$, in which the WD can increase its mass to $1.378{M}_\odot$ and produce SN Ia. The parameter space obtained here is smaller than that of Wang et al. (2009a). This is because for lower values of $\eta_{\rm He}$, some WDs in binary systems cannot increase their masses to the Ch-mass limit by He nova outbursts. In Fig.\,9, we present regions in the orbital period-companion mass plane for WD binaries that can produce SNe Ia for initial WD masses of 0.9, 1.0, 1.1, and $1.2{M}_\odot$. We found that the region almost vanishes if the ${M}_{\rm WD}^{\rm i}<0.9{M}_\odot$, which is then assumed to be the minimum WD mass for producing SNe Ia from the WD + He star channel.

\begin{figure}[]
\begin{center}
\includegraphics[width=9.5cm,angle=0]{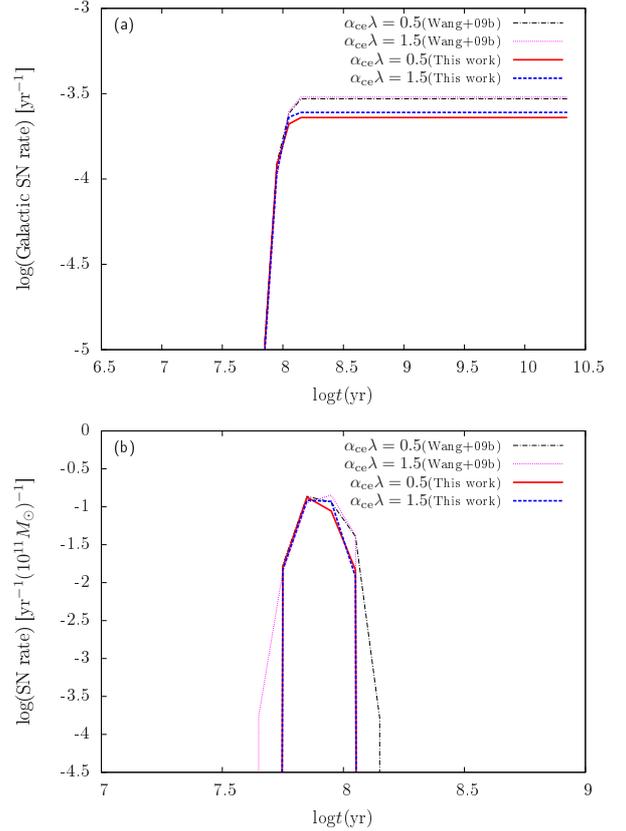}
 \caption{Panel (a): Evolution of Galactic birthrates of SNe Ia for a constant star formation rate (SFR=$5\,{M}_\odot\,\mbox{yr}^{-1}$). The red and blue lines are results from our work, whereas the black and pink lines are taken from Wang et al. (2009b). Panel (b): Same as panel (a), but for a single starburst with a total mass of ${10}^{11}{M}_\odot$.}
  \end{center}
\end{figure}

The mass retention efficiencies of He nova outbursts given in our simulations are lower than KH04, which may have influence on the SN Ia birthrates for the WD + He star channel. In order to investigate the influence, we performed a series of Monte Carlo simulations in the BPS study using the Hurley rapid binary evolution code (see Hurley et al. 2002). The methods and assumptions in our simulations are similar to Wang et al. (2009b). In Fig.\,10, panel (a) presents the evolution of Galactic birthrates of SNe Ia for a constant star formation. The SNe Ia birthrates in the Galaxy given in our work range from $2.3\times{10}^{-4}$ to $2.5\times{10}^{-4}\,\mbox{yr}^{-1}$, which are lower than those of Wang et al. (2009b). This is due to the smaller parameter space obtained in our work. Panel (b) shows the evolution of SN Ia birthrates for a single starburst with a total mass of ${10}^{11}{M}_\odot$. The range of delay time distribution is between $5.6\times{10}^{7}$ to $1.1\times{10}^{8}\,\mbox{yr}$ after the starburst, which can explain SNe Ia with short delay times.

\section{Discussion}

\begin{figure}[]
\begin{center}
\includegraphics[width=9.5cm,angle=0]{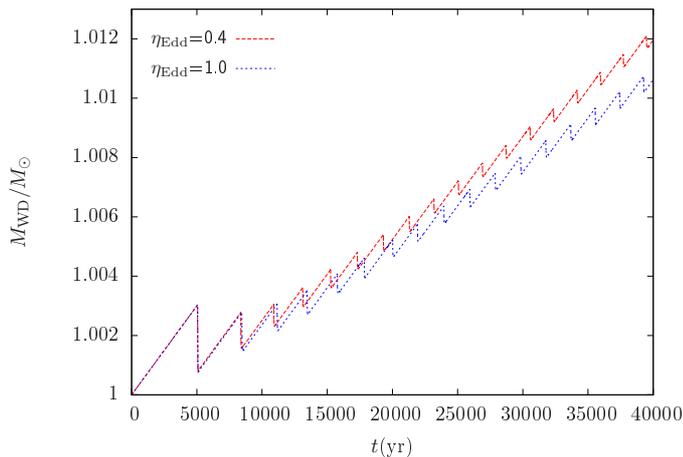}
 \caption{Mass of the WD as a function of time during the process of mass accumulation under different super-Eddington wind parameters ($\eta_{\rm Edd}$), where the initial WD mass and accretion rate are $1.0{M}_\odot$ and $6\times10^{-7}\,{M}_\odot\,\mbox{yr}^{-1}$, respectively.}
  \end{center}
\end{figure}

In our simulations, the super-Eddington wind is assumed to be the mass-loss mechanism during He-layer flashes, where the super-Eddington wind factor $\eta_{\rm Edd}$ used in our simulations is set to be $1$, which means that the super-Eddington wind has the maximum ejection efficiency. However, $\eta_{\rm Edd}$ is more likely lower than $1$ owing to the energy dissipation of absorption and ionization processes. Thus, the mass retention efficiencies $\eta_{\rm He}$ in our simulations should be the lower limit under the super-Eddington wind assumption. To investigate the influence of $\eta_{\rm Edd}$ on the process of mass accumulation, we show the WD mass as a function of time under two different values of $\eta_{\rm Edd}$ in Fig.\,11. The initial WD mass and accretion rate are $1.0{M}_\odot$ and $6\times10^{-7}\,{M}_\odot\,\mbox{yr}^{-1}$, respectively. The blue dashed line is the mass variation under the $\eta_{\rm Edd}=1$. For comparison, the red dashed line presents the mass variation under the $\eta_{\rm Edd}=0.4$, which means that only $40\%$ of the redundant energy is used to offer the kinetic energy of the ejecta when the ${L}_{\rm eff}$ exceeds $L_{\rm Edd}$. From this figure, we can see that $\eta_{\rm He}$ increases with the decline of $\eta_{\rm Edd}$. This is because the super-Eddington wind becomes weaker for lower $\eta_{\rm Edd}$. Furthermore, we have not considered the influence of rotation on the process of accretion; the He burning is much less violent if the rotation of WD is taken into account, which may increase the value of $\eta_{\rm He}$ (e.g., Yoon et al. 2004).

\begin{figure}[]
\begin{center}
\includegraphics[width=9.3cm,angle=0]{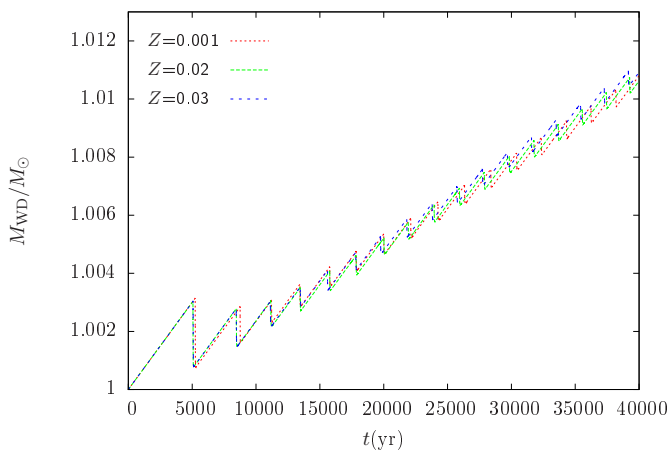}
 \caption{Similar to Fig.\,11, but under various metallicities of accreted He-rich material.}
  \end{center}
\end{figure}

Previous studies suggested that metallicities may have influence on the mass accumulation processes and the maximum luminosities of SNe Ia (e.g., Timmes et al. 2003; Podsiadlowski et al. 2008; Meng et al. 2009). In Fig.\,12, we present the WD mass as a function of time during the process of mass accumulation under various metallicities of accreted He-rich material, where the initial WD mass and accretion rate in our calculations are set to be $1.0{M}_\odot$ and $6\times10^{-7}\,{M}_\odot\,\mbox{yr}^{-1}$, respectively. If the accreted He-rich material has a high value of metallicity, the opacity on the surface of the WD will be increased, resulting in that the super-Eddington wind can be triggered more easily. This indicates that the He novae at high metallicities have shorter nova cycle durations. Moreover, the super-Eddington wind can even be triggered at extremely low metallicity (e.g., $Z=10^{-6}$; see Ma et al. 2013), which implies that the super-Eddington wind scenario may contribute toward explaining the SNe Ia at high redshifts. Although the metallicities have influence on the opacity and thus the processes of wind mass loss, the opacity is more sensitive to the temperature and density of the material, which means that metallicities of accreted material may not be the principal parameter that influences mass retention efficiencies. Thus, the differences among various metallicities shown in this simulation are not obvious.

Recurrent novae are those whose outbursts occur more than once in a century (e.g., Maoz et al. 2014). Hillman et al. (2016) have investigated hydrogen flashes and found that the cycle durations and flash durations can be used to derive the key parameters of recurrent novae. The He novae studied in our simulations have longer cycle durations than classical novae. This is because hydrogen is more easily ignited than He on the surface of a WD, which means that the hydrogen-rich layer has less mass than the He-rich layer when nova outbursts occur. Therefore, classical nova outbursts are more frequent; this corresponds to the shorter cycle durations for the same initial WD mass and accretion rates.

The accretion rates in our simulations range from $4\times10^{-8}\,{M}_\odot\,\mbox{yr}^{-1}$ to $1.6\times10^{-6}\,{M}_\odot\,\mbox{yr}^{-1}$, and multicycle He-layer flashes can be triggered under these accretion rates. For a WD, if the accretion rate is higher than its critical rate for stable He burning, the He flashes will become too weak to eject mass and the accumulated He-rich material can be burned into carbon and oxygen steadily. In contrast, if the accretion rate is lower than $4\times10^{-8}\,{M}_\odot\,\mbox{yr}^{-1}$, the accumulated He-rich material can be degenerate on the surface of the WD and the He detonation should be triggered when the mass of the He layer reaches a critical value (e.g., Woosley \& Weaver 1986). The He detonation wave propagates inward, compressing the CO core and triggers another detonation wave propagating from the inside out; this is called the double detonation explosion model (e.g., Taam 1980; Nomoto 1982a; H\"oflich \& Khokhlov 1996; H\"oflich et al. 2013). Wang et al. (2013) have suggested that the double detonation model may explain the formation of type Iax SNe; the surviving companions from this model may produce a hypervelocity He star, such as US 708 (e.g., Geier et al. 2015; Neunteufel et al. 2016). However, it is still not clear how the He layer detonation is triggered, which may lead to the existence of some uncertainties about the final outcomes of WD (e.g., Sim et al. 2010; Fink et al. 2010).

\section{Summary}

By employing the stellar evolution code MESA, we investigated the mass retention efficiencies of CO WDs that undergo multicycle nova outbursts via accretion processes. We obtained the mass retention efficiencies of He nova outbursts for various initial WDs that can be used in binary population synthesis studies. We also obtained the nova cycle durations for various WDs and found that the durations of He novae are much longer than those of the hydrogen novae. Meanwhile, we found that a WD can increase its mass to the Ch-mass limit if nova outbursts continue to occur, trigger the explosive carbon burning in its center and finally exploding as SN Ia. These results indicate that SNe Ia can be produced by multicycle He nova outbursts. Furthermore, the elemental abundance profile at the moment of explosion obtained in our simulations can be used to provide initial input parameters for explosion models of SNe Ia. Moreover, we systematically simulated binary evolution of WD + He star systems and obtained the parameter spaces for producing SNe Ia from this channel. The BPS results revealed that the birthrates of SNe Ia are lower than those in Wang et al. (2009b), which is due to the lower mass retention efficiencies obtained in this work. In the future, we hope that more He novae can be identified in the observations, which will be useful for the theoretical studies of He-layer flashes and the progenitors of SNe Ia.

\begin{acknowledgements}

We acknowledge useful comments and suggestions from the referee. This work is supported by the National Basic Research Program of China (973 programme, 2014CB845700), the National Natural Science Foundation of China (Nos 11673059, 11521303, 11390374, 11573016 and 11547041), the Chinese Academy of Sciences (Nos KJZD-EW-M06-01), the Key Research Program of Frontier Science CAS (No QYZDB-SSW-SYS001), the Youth Innovation Promotion Association CAS, and the Natural Science Foundation of Yunnan Province (Nos 2013HB097 and 2013HA005).

\end{acknowledgements}

\appendix
\section{Fitting formulae for mass retention efficiencies}
In this appendix, we have provided the fitting formula of mass retention efficiencies ($\eta_{\rm He}$) for different initial WD models that can be used in the binary population synthesis studies. The $\eta_{\rm He}$ for various WDs can be expressed as a function of accretion rate ($\log\dot{M}_{\rm acc}$) by the following relation:
\begin{equation}
\eta_{\rm He}=a+b\times\log\dot{M}_{\rm acc}+c\times(\log\dot{M}_{\rm acc})^2,
\end{equation}
where $a$, $b$, and $c$ are coefficients in this formula, which are given in the Table A.1.

\begin{table*}
\centering
\caption{Fitting formula coefficients of $\eta_{\rm He}$ for various WDs, where $x$ in the table represents the accretion rate ($\log\dot{M}_{\rm acc}$).}
\begin{tabular}{|c|c|c|c|c|}     
\hline
 ${M}_{\rm WD}$ & $\log\dot{M}_{\rm acc}$& $a$  &$b$  &$c$\\
 \hline
 0.7  & $-6.52\leq{x}$    & $1$        & $0$       & $0$\\
      & $-6.66<x<-6.52$ & $1485.13$  & $445.86$  & $33.47$\\
      & $-6.75<x<-6.66$ & $79.96$    & $20.91$   & $1.50$\\
      & $-6.92<x<-6.75$ & $-30.19$   & $-9.09$   & $-0.70$\\
 \hline
 0.8  & $-6.38\leq{x}$    & $1$        & $0$       & $0$\\
      & $-6.50<x<-6.38$ & $2135.61$  & $658.46$  & $50.76$\\
 \hline
 0.9  & $-6.27\leq{x}$    & $1$        & $0$       & $0$\\
      & $-6.59<x<-6.27$ & $292.58$   & $89.22$   & $6.81$\\
 \hline
 1.0  & $-6.10\leq{x}$    & $1$        & $0$       & $0$\\
      & $-6.16<x<-6.10$ & $51.67$    & $8.31$    & $0$\\
      & $-7.05<x<-6.61$ & $21.66$    & $6.08$    & $0.43$\\
 \hline
 1.1  & $-5.89\leq{x}$    & $1$        & $0$       & $0$\\
      & $-6.10<x<-5.89$  &$597.09$   & $197.31$  & $16.31$\\
      & $-7.10<x<-6.10$ & $25.45$    & $7.23$    & $0.52$\\
 \hline
 1.2  & $-5.82\leq{x}$    & $1$        & $0$       & $0$\\
      & $-7.22<x<-5.82$ & $25.19$    & $7.15$    & $0.51$\\
 \hline
 1.25 & $-5.80\leq{x}$    & $1$        & $0$       & $0$\\
      & $-7.22<x<-5.80$ & $22.57$    & $6.34$    & $0.45$\\
 \hline
 1.3  & $-5.80\leq{x}$    & $1$        & $0$       & $0$\\
      & $-7.22<x<-5.80$ & $18.52$    & $5.05$    & $0.35$\\
 \hline
 1.35 & $-5.80\leq{x}$    & $1$        & $0$       & $0$\\
      & $-5.82<x<-5.80$ & $28.50$    & $4.74$    & $0$\\
      & $-7.30<x<-5.82$ & $11.31$    & $2.81$    & $0.18$\\
 \hline
\end{tabular}
\end{table*}

\section{Fitting formulae for nova cycle durations}
This appendix provides fitting formulae of nova cycle durations $D$ (in years) as a function of $\log\dot{M}_{\rm acc}$ (in ${M}_\odot\,\mbox{yr}^{-1}$) in linear form as follows:
\begin{equation}
\log D=a-b\times10^{10}\,\log\dot{M}_{\rm acc},
\end{equation}
where the coefficients $a$ and $b$ are given in Table B.1.

\begin{table*}
\centering
\caption{Fitting formula coefficients of nova cycle durations.}
\begin{tabular}{|c|c|c|}     
\hline
 ${M}_{\rm WD}$ & $a$  & $b$\\
 \hline
 0.8  & 62311.80    & 13.3\\
 \hline
 0.9  & 40691       & 7.62\\
 \hline
 1.0  & 46352.24    & 8\\
 \hline
 1.1  & 25404.53    & 2.80\\
 \hline
 1.2  & 15511.62    & 1.51\\
 \hline
 1.25 & 11399.37    & 1.0\\
 \hline
 1.3  & 5240.06     & 0.455\\
 \hline
 1.35 & 2211.85     & 0.193\\
 \hline
\end{tabular}
\end{table*}


\begin{thebibliography}{}
\bibitem[Ashok \& Banerjee (2003)]{ash03}        Ashok, N. M., \& Banerjee, D. P. K. 2003, A\&A, 409, 1007
\bibitem[Bildsten et al. (2007)]{Bild07}         Bildsten, L., Shen, K. J., Weinberg, N. N., \& Nelemans, G. 2007, ApJL, 662, L95
\bibitem[Brooks et al. (2016)]{Bro16}            Brooks, J., Bildsten, L., Schwab, J., \& Paxton, B. 2016, ApJ, 821, 28
\bibitem[Chen et al. (2014)]{Chen14}             Chen, X., Han, Z., \& Meng, X. 2014, MNRAS, 438, 3358
\bibitem[Denissenkov et al. (2013)]{Den13}       Denissenkov, P. A., Herwig, F., Bildsten, L., \& Paxton, B. 2013, ApJ, 762, 8
\bibitem[Denissenkov et al. (2017)]{Den17}       Denissenkov, P. A., Herwig, F., Battino, U., Ritter, C., Pignatari, M., et al. 2017, ApJL, 834, L10
\bibitem[Eggleton (1973)]{Eg73}                  Eggleton, p. p. 1973, MNRAS, 163, 279
\bibitem[Fink et al. (2010)]{Fink10}             Fink, M., R\"opke, F. K., Hillebrandt, W., et al. 2010, A\&A, 514, 53
\bibitem[Geier et al. (2015)]{Ger15}             Geier, S., F\"urst, F., Ziegerer, E., Kupfer, T., Heber, U., et al. 2015, Science, 347, 1126
\bibitem[Hachisu et al. (1996)]{Ha96}            Hachisu, I., Kato, M., \& Nomoto, K. 1996, ApJL, 470, L97
\bibitem[Han \& Podsiadlowski (2004)]{han04}     Han, Z., \& Podsiadlowski, P. 2004, MNRAS, 350, 1301
\bibitem[Hillman et al. (2015)]{Hill05}          Hillman, Y., Prialnik, D., Kovetz, A., \& Shara, M. M. 2015, MNRAS, 446, 1924
\bibitem[Hillman et al. (2016)]{Hill06}          Hillman, Y., Prialnik, D., Kovetz, A., \& Shara, M. M. 2016, ApJ, 819, 168
\bibitem[H\"oflich et al. (1995)]{Hoef95}        H\"oflich, P., Khokhlov, A. M., \& Wheeler, J. C. 1995, ApJ, 444, 831
\bibitem[H\"oflich \& Khokhlov (1996)]{HK96}     H\"oflich, P., \& Khokhlov, A. M. 1996, ApJ, 457, 500
\bibitem[H\"oflich et al. (2013)]{Hoef13}        H\"oflich, P., Dragulin, P., Mitchell, J., et al. 2013, Frontiers of Physics, 8, 144
\bibitem[Hoyle \& Fowler (1960)]{HF60}           Hoyle, F., \& Fowler, W. A. 1960, ApJ, 132, 565
\bibitem[Hurley et al. (2002)]{Hu02}             Hurley, J. R., Tout, C. A., \& Pols, O. R. 2002, MNRAS, 329, 897
\bibitem[Iben \& Tutukov (1984)]{IT84}           Iben, I., \& Tutukov, A. V. 1984, ApJS, 54, 335
\bibitem[Igleslas \& Rooers (1996)]{IR96}        Igleslas, C. A., \& Rooers, F. J. 1996, ApJ, 464, 943
\bibitem[Jones et al. (2013)]{Jones13}           Jones, D. O., Rodney, S. A., Riess, A. G., et al. 2013, ApJ, 768, 166
\bibitem[Kato \& Hachisu (2004)]{KH04}           Kato, M., \& Hachisu, I. 2004, ApJL, 613, L129 (KH04)
\bibitem[Kato et al. (2008)]{KH08}               Kato, M., Hachisu, I., Kiyota, S., \& Saio, H. 2008, ApJ, 684, 1366
\bibitem[Khokhlov (1991)]{Kh91}                  Khokhlov, A. M. 1991, A\&A, 245, 114
\bibitem[Kobayashi et al. (1998)]{Ko98}          Kobayashi, C., Tsujimoto, T., Nomoto, K., Hachisu, I., \& Kato, M. 1998, ApJL, 503, L155
\bibitem[Li \& van den Heuvel (1997)]{Li97}      Li, X.-D., \& van den Heuvel, E. P. J. 1997, A\&A, 322, L9
\bibitem[Ma et al. (2013)]{ma13}                 Ma, X., Chen, X., Chen, H., Denissenkov, P., \& Han, Z. 2013, ApJL, 778, L32
\bibitem[Maoz et al. (2014)]{mao14}              Maoz, D., Mannucci, F., \& Nelemans, G. 2014, ARA\&A, 52, 107
\bibitem[Meng et al. (2009)]{Meng09}             Meng, X., Chen, X., \& Han, Z. 2009, MNRAS, 395, 2103
\bibitem[Nelemans et al. (2001)]{nel01}          Nelemans, G., Portegies Zwart, S. F., Verbunt, F., \& Yungelson, L. R. 2001, A\&A, 368, 939
\bibitem[Neunteufel et al. (2016)]{Neu16}        Neunteufel, P., Yoon, S.-C., \& Langer, N. 2016, A\&A, 589, A43
\bibitem[Nomoto (1982a)]{Nom82a}                 Nomoto, K. 1982a, ApJ, 257, 780
\bibitem[Nomoto (1982b)]{Nom82b}                 Nomoto, K. 1982b, ApJ, 253, 798
\bibitem[Nomoto et al. (1984)]{Nom84}            Nomoto, K., Thielemann, F. K. \& Yokoi, K. 1984, ApJ, 286, 644
\bibitem[Nomoto et al. (1997)]{Nom97}            Nomoto, K., Iwamoto, K., \& Kishimoto, N. 1997, Science, 276, 1378
\bibitem[Paxton et al. (2011)]{pax11}            Paxton, B., Bildsten, L., Dotter, A., et al. 2011, ApJS, 192, 3
\bibitem[Paxton et al. (2013)]{pax13}            Paxton, B., Cantiello, M., Arras, Ph., et al. 2013, ApJS, 208, 4
\bibitem[Paxton et al. (2015)]{pax15}            Paxton, B., Marchant, P., Schwab, J., et al. 2015, ApJS, 220, 15
\bibitem[Piersanti et al. (2014)]{Pi14}          Piersanti, L., Tornamb\'e, A., \& Yungelson, L. R. 2014, MNRAS, 445, 3239
\bibitem[Piersanti et al. (2015)]{pier15}        Piersanti, L., Tornamb\'{e}, A., \& Yungelson, L. R. 2015, MNRAS, 452, 2897
\bibitem[Podsiadlowski et al. (2008)]{Pod08}     Podsiadlowski, P., Mazzali, P., Lesaffre, P., Han, Z., \& F\"orster, F. 2008, New Astron. Rev., 52, 381
\bibitem[Ruiter et al. (2009)]{Ru09}             Ruiter, A. J., Belczynski, K., \& Fayer, C. 2009, ApJ, 699, 2026
\bibitem[Shen \& Bildsten (2009)]{Shen09}        Shen, K. J., \& Bildsten, L. 2009, ApJ, 699, 1365
\bibitem[Sim et al. (2010)]{Sim10}               Sim, S. A., R\"opke, F. K., Hillebrandt, W., et al. 2010, ApJL, 714, L52
\bibitem[Taam (1980)]{Ta80}                      Taam, R. E. 1980, ApJ, 237, 142
\bibitem[Timmes et al. (2003)]{Tim03}            Timmes, F. X., Brown, E. F., \& Truran, J. W. 2003, ApJ, 590, L83
\bibitem[Wang et al. (2009a)]{Wang09a}           Wang, B., Meng, X., Chen, X., \& Han, Z. 2009a, MNRAS, 395, 847
\bibitem[Wang et al. (2009b)]{Wang09b}           Wang, B., Chen, X., Meng, X., \& Han, Z. 2009b, ApJ, 701, 1540
\bibitem[Wang \& Han (2012)]{Wang12}             Wang, B., \& Han, Z. 2012, New Astron. Rev., 56, 122
\bibitem[Wang et al. (2013)]{Wang13}             Wang, B., Justham, S., \& Han, Z. 2013, A\&A, 559, A94
\bibitem[Wang et al. (2015)]{Wang15}             Wang, B., Li, Y., Ma, X., et al. 2015, A\&A, 584, A37
\bibitem[Webbink (1984)]{Web84}                  Webbink, R. F. 1984, ApJ, 277, 355
\bibitem[Whelan \& Iben (1973)]{WI73}            Whelan, J., \& Iben, I. 1973, ApJ, 186, 1007
\bibitem[Woosley \& Weaver (1986)]{WW86}         Woosley, S. E., \& Weaver, T. A. 1986, ARA\&A, 24, 205
\bibitem[Woudt et al. (2009)]{wou09}             Woudt, P. A., Steeghs, D., Karovska, M., et al. 2009, ApJ, 706, 738
\bibitem[Wu et al. (2016)]{Wu16}                 Wu, C., Liu, D., Zhou, W., \& Wang, B. 2016, Res. Astron. Astrophys., 16, 160
\bibitem[Yoon et al. (2004)]{Yoon04}             Yoon, S.-C., Langer, N., \& Scheithauer, S. 2004, A\&A, 425, 217

\end{thebibliography}
\end{document}